\documentclass[fleqn,twoside]{article}
\usepackage{espcrc2}
\usepackage[]{epsfig}
\usepackage{amssymb}
\usepackage{afterpage}

\newcommand{\Tr}{\mbox{Tr}}
\newcommand{\Or}{\mbox{${\cal O}$}}

\title{\vspace{-1.25cm}
  {\normalsize DESY 05-248}\\[-0.2cm]
  {\normalsize HU-EP-05/77} \\[0.50cm]
  Probing the topological structure of the QCD vacuum with overlap fermions}

\author{E.-M.~Ilgenfritz\address{Institut f\"ur Physik, Humboldt Universit\"at zu Berlin, 12489 Berlin, Germany}, K.~Koller\address{Sektion Physik, Universit\"at M\"unchen, 80333 M\"unchen, Germany}, Y.~Koma\address[DESY]{Deutsches Elektronen-Synchrotron DESY, 22603 Hamburg, Germany}, G.~Schierholz$^{\rm c,}$\address[NIC]{John von Neumann-Institut f\"ur Computing NIC, 15738 Zeuthen, Germany}, T.~Streuer\addressmark[NIC], V.~Weinberg$^{\rm d,}$\address{Institut f\"ur theoretische Physik, Freie Universit\"at Berlin, 14196 Berlin, Germany}\thanks{Speaker}}

\begin{document}

\begin{abstract}
Overlap fermions implement exact chiral symmetry on the lattice and are thus an appropriate tool for investigating the chiral and topological structure of the QCD vacuum.
We study various chiral and topological aspects on L\"uscher-Weisz-type quenched  gauge field configurations using overlap fermions as a probe.
Particular emphasis is placed upon the analysis of the spectral density and the localisation properties of the eigenmodes  as well as on the local structure of topological charge fluctuations. 

\vspace{1pc}
\end{abstract}

\maketitle

\section{Introduction}
\thispagestyle{empty}

The understanding of the vacuum structure of QCD is an important goal of elementary particle physics. The distribution of topological charge density is believed to describe nonperturbative phenomena like the large $\eta'$ mass, the axial U(1) anomaly, the $\theta$ dependence and the spontaneous breaking of chiral symmetry.
In the traditional instanton picture the excitations of the QCD vacuum are  phenomenologically modelled as an interacting ensemble of instantons and anti-instantons.

Recent advances in implementing chiral symmetry on the lattice have made it possible to directly clarify from  first principles to what extent the instanton picture of localised solutions of the Euclidean Yang-Mills equations of motion realistically describes the QCD vacuum.

Overlap fermions~\cite{Neuberger:1997fp,Neuberger:1998wv}
have an exact chiral symmetry on the lattice~\cite{Luscher:1998pq} and are the cleanest known theoretical description of lattice fermions.
Their implementation of chiral symmetry and the possibility to exactly define the index theorem on the lattice allow us to investigate the relationship of topological properties of gauge fields and the dynamics of fermions.
A further advantage of overlap fermions, in contrast to Wilson fermions, is that they are automatically $\Or(a)$ improved~\cite{Capitani:1999uz} and not plagued by exceptional configurations.

The  overlap operator is defined by
\begin{equation}
D_N=\frac{\rho}{a}\big(1+\frac{X}{\sqrt{X^\dagger X}}\big),
X=D_W-\frac{\rho}{a},
\label{DN}
\end{equation}
where we use the Wilson-Dirac operator $D_W$ as the kernel of the overlap operator.

The operator $D_N$ has $n_- + n_+$ exact zero modes, $D_N \psi_n = 0$, $n_-$ ($n_+$) being the number of modes 
with negative (positive) chirality, $\gamma_5 \psi_n = - \psi_n$ ($\gamma_5 
\psi_n = + \psi_n$).  The index of $D_N$ is  given by $Q = n_- - n_+$.
The nonzero modes $\lambda$ with $D_N \psi_\lambda = \lambda \psi_\lambda$  come in complex 
conjugate pairs $\lambda$ and $\lambda^*$ and satisfy $(\psi_\lambda^\dagger,\gamma_5 \psi_\lambda^{}) = 0$.

The mass parameter $\rho$  is chosen to be $1.4$, which is a compromise between the physical
requirement of good locality properties of the overlap operator and the performance requirement of a small condition number of $D_W$.

To compute the `sign function'
\begin{equation}
{\rm sgn}(X) = \frac{X}{\sqrt{X^\dagger X}} \equiv 
\gamma_5\, {\rm sgn}(H),\, H = \gamma_5 X, 
\end{equation}
we use the minmax polynomial approximation~\cite{Giusti:2002sm}.
We compute the ${\cal O}(10)$ lowest eigenvalues of $H$ and treat the sign 
function on the corresponding subspace exactly.

Topological studies using the Wilson gauge field action suffer from dislocations \cite{Gockeler:1989qg} and should be treated with caution. We therefore use the L\"uscher-Weisz gauge field action \cite{Luscher:1984xn}, which suppresses dislocations and greatly reduces the number of unphysical zero modes.

The L\"uscher-Weisz gauge field action is given by
\begin{eqnarray}
S[U]&\!\!\!=\!\!\!&\frac{6}{g^2}\Big[ c_0\!\!\!\! 
\sum_{\rm plaquette}\frac{1}{3}\,
\mbox{Re}\, 
\mbox{Tr}\, (1-U_{\rm plaquette}) \nonumber \\
&\!\!\!+\!\!\!& c_1\!\!\!\! \sum_{\rm rectangle}\frac{1}{3}\, \mbox{Re}\, 
\mbox{Tr}\, (1-U_{\rm rectangle}) \label{ImpAct} \\
&\!\!\!+\!\!\!& c_2\!\!\!\!\!\!\!\! \sum_{\rm parallelogram}\frac{1}{3}\, 
\mbox{Re}\, \mbox{Tr}\, (1-U_{\rm parallelogram}) \Big], \nonumber
\end{eqnarray}
with coefficients $c_1$, $c_2$ ($c_0 + 8 c_1 + 8 c_2 = 1$) taken from tadpole 
improved perturbation theory~\cite{Gattringer:2001jf}.

To investigate the volume dependence of our data we have simulated at 3 different volumes at fixed coupling $\beta=8.45$. 
To explore the $a$ dependence of the results, we also employed a  $12^3\times 24$ lattice at $\beta=8.10$ with approximately the same physical volume as the $16^3\times 32$ lattice.

In the following table we list the lattices used in our  calculations, where we take $r_0=$0.5 fm to set the scale:

\vspace*{1cm}

\begin{tabular}{|c|c|c|c|c|c|}
    \hline
    $\beta$ & $a$ [fm] & V            & confs  & modes\\\hline\hline
    8.45    & 0.095    & $12^3\times 24$  & 500       & {\Or}(50)\\
    8.45    & 0.095    & $16^3\times 32$ & 267       & {\Or}(140)\\
    8.45    & 0.095    & $24^3\times 48$ & 186       & {\Or}(160)\\\hline
    8.10    & 0.125    & $12^3\times 24$ & 254       & {\Or}(140)\\
    \hline
\end{tabular}

\vspace*{1cm}

\section{Density and locality of the eigenmodes of the overlap operator}

The spontaneous breaking of chiral symmetry by the dynamical creation of a 
nonvanishing chiral condensate, 
$\langle\bar\Psi\Psi\rangle$, 
is related to the spectral density 
$\rho(\lambda)$ of the Dirac operator near zero by the 
Banks-Casher relation \cite{Banks:1979yr} 
$\langle\bar\Psi\Psi\rangle=- \pi/V \rho(0)$. 

The eigenvalues of $D_N$ lie on a circle around ($\rho$,0) with radius
$\rho$ in the complex plane. In our spectral analysis we use the improved operator 
$D_N^{\rm imp} = (1-aD_N/2\rho)^{-1} D_N$ 
\cite{Chiu:1998gp,Chiu:1998aa}
, which projects the
eigenvalues of $D_N$ stereographically onto the imaginary axis.
The nonzero eigenvalues of $D_N^{\rm imp}$ come in pairs $\pm {\rm i}
\bar\lambda$, while the zero modes are untouched.   

Using the Arnoldi-algorithm, we compute the $\Or(50)-\Or(160)$ lowest eigenvalues on every configuration.

The spectral density of the `continuous' modes is formally given by
\begin{equation}
\rho(\lambda)=\frac{1}{V} \big\langle\sum_{\bar{\lambda}} \delta(\lambda
-\bar{\lambda})\big\rangle ,
\end{equation}
where the sum extends over the nonzero eigenvalues $\pm {\rm i}
\bar{\lambda}$ of $D_N^{\rm imp}$. 

\begin{figure}[h]
\begin{center}
\hspace*{-0.6cm}\epsfig{file=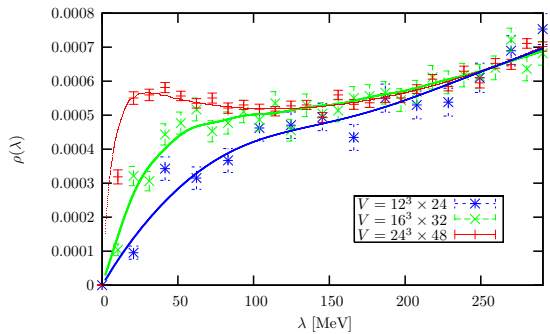, width=6.5cm}
\end{center}
\caption{The spectral density at $\beta=8.45$ together with a fit using the finite volume predictions of quenched chiral perturbation theory.}
\label{fig:density}
\end{figure}

Fig.~\ref{fig:density} shows the spectral density together with a fit\footnote{
To consider the effects of higher orders in chiral perturbation theory we added a term $a_1 \lambda + a_2 \lambda^2$ to the fitted formula (\ref{rho}).}
 using the prediction of chiral perturbation theory.
For small eigenvalues one can see a strong volume dependence of the spectral density.

In a finite volume, the spectral density is given in quenched chiral perturbation theory \cite{Osborn:1998qb,Damgaard:2001xr} as 

\begin{equation}
\rho(\lambda)=\Sigma_{\rm eff}(V,\lambda)\sum_Q w(Q) \rho_Q(V\Sigma_{\rm eff}(V,\lambda)\lambda) ,
\label{rho}
\end{equation}

\noindent where $\rho_Q(x)=\frac{x}{2}(J_{|Q|}^2(x)-J_{|Q|+1}(x)J_{|Q|-1}(x))$ is the microscopic spectral density \cite{Wilke:1997gf} in the sector of fixed topological charge $Q$ defined in terms of Bessel functions $J_n(x)$.

The effective chiral condensate is given by $\Sigma_{\rm eff}(V,\lambda)=\Sigma(1-16\pi^2\delta \bar{G}'_V(2\Sigma\lambda/f^2))$, where $\delta$ is the chiral log parameter, $f$ denotes the pion decay constant  and  $\bar{G}'_V(M^2)=\partial_{M^2}\bar{G}_V(M^2)$ is the derivative of the pseudo-Goldstone boson propagator without zero momentum modes $\bar{G}_V(M^2)=1/V\sum_p (p^2+M^2)^{-1}-(M^2 V)^{-1}$. Taking  the weight $w(Q)$ of the sector of topological charge $Q$ with
$\sum_Q w(Q) = 1$  from our `measured' charge distributions,
we obtain $\delta=0.20(3)$ and  $\Sigma=(212(18)\, \mbox{MeV})^3$ ($\Sigma^{\overline{\mbox{\tiny MS}}}(\mbox{2\,GeV})=(231(20)\, \mbox{MeV})^3)$.

A useful measure  to quantify the localisation of eigenmodes \cite{Gattringer:2001mn,Aubin:2004mp,Gubarev:2005jm} is the inverse participation ratio (IPR)
\begin{equation}
I=V\sum_x\rho(x)^2, 
\end{equation}
with the scalar density 
$\rho(x)=\psi_\lambda^\dagger(x)\psi_\lambda(x)$ for
normalised eigenfunctions $\sum_x\rho(x)=1$. 
While $I=V$ if the density has support only on one lattice point, 
$I$ decreases as the density becomes more delocalised, 
reaching $I=1$ when the density is maximally spread over all lattice sites. 

\begin{figure*}[t]
\begin{center}
\begin{tabular}{cc}
\epsfig{file=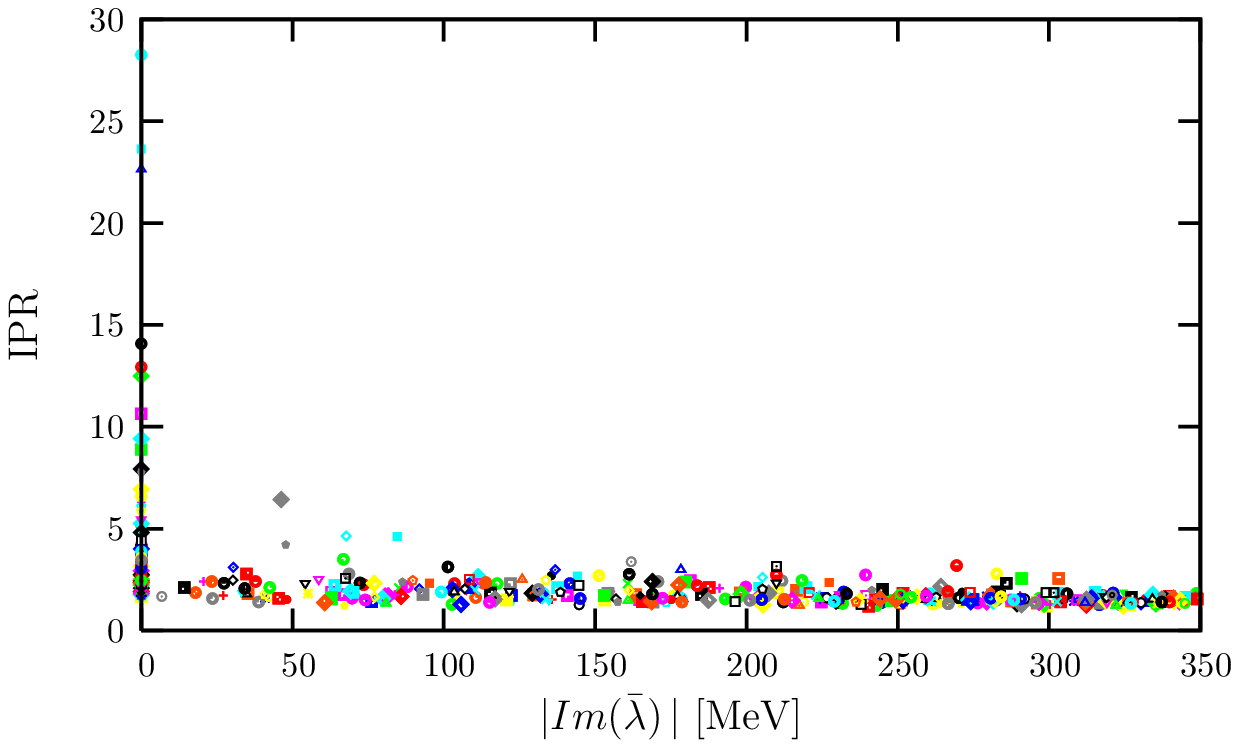,width=6cm}&\epsfig{file=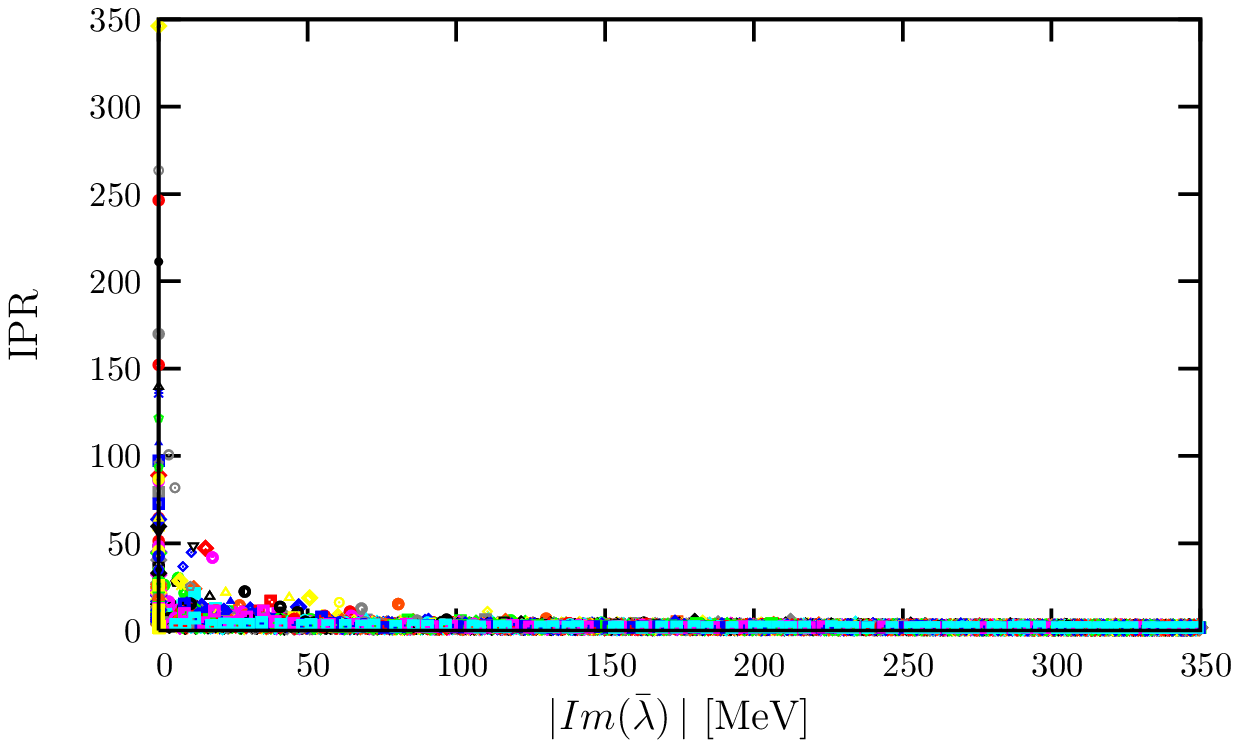,width=6cm}\\
(a) & (b)\\
\epsfig{file=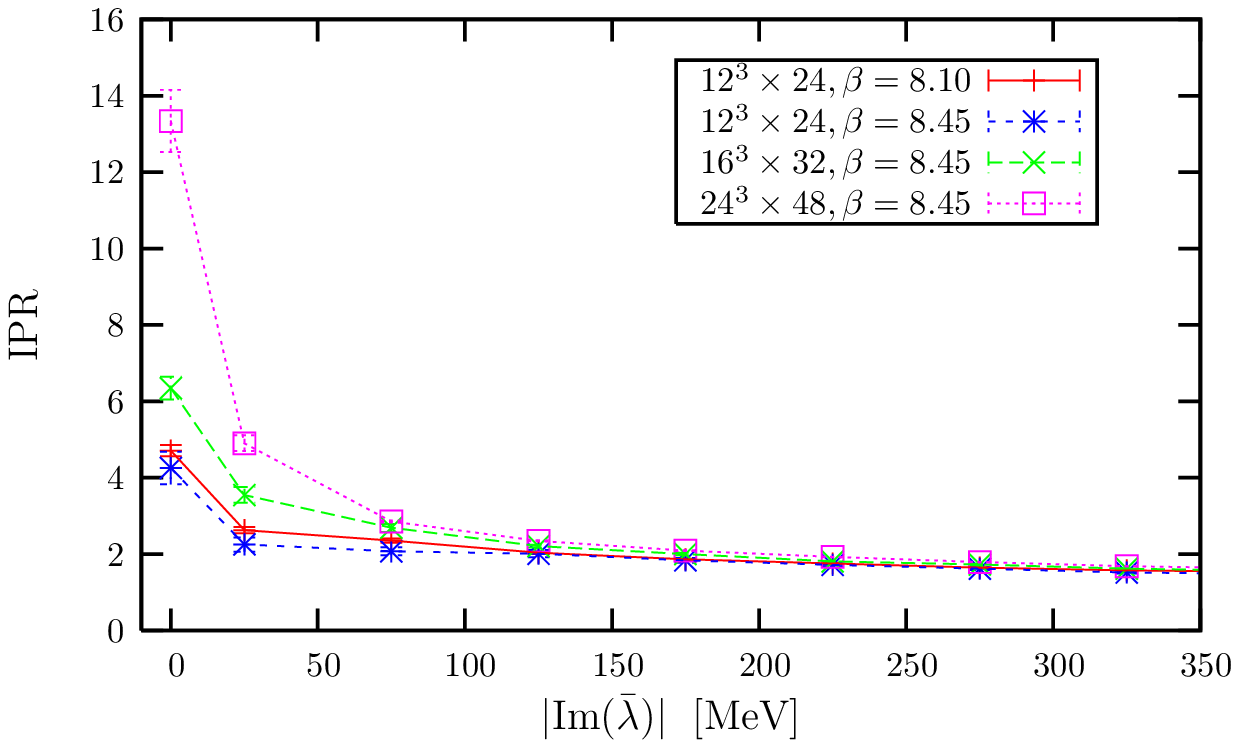,width=6cm}&\epsfig{file=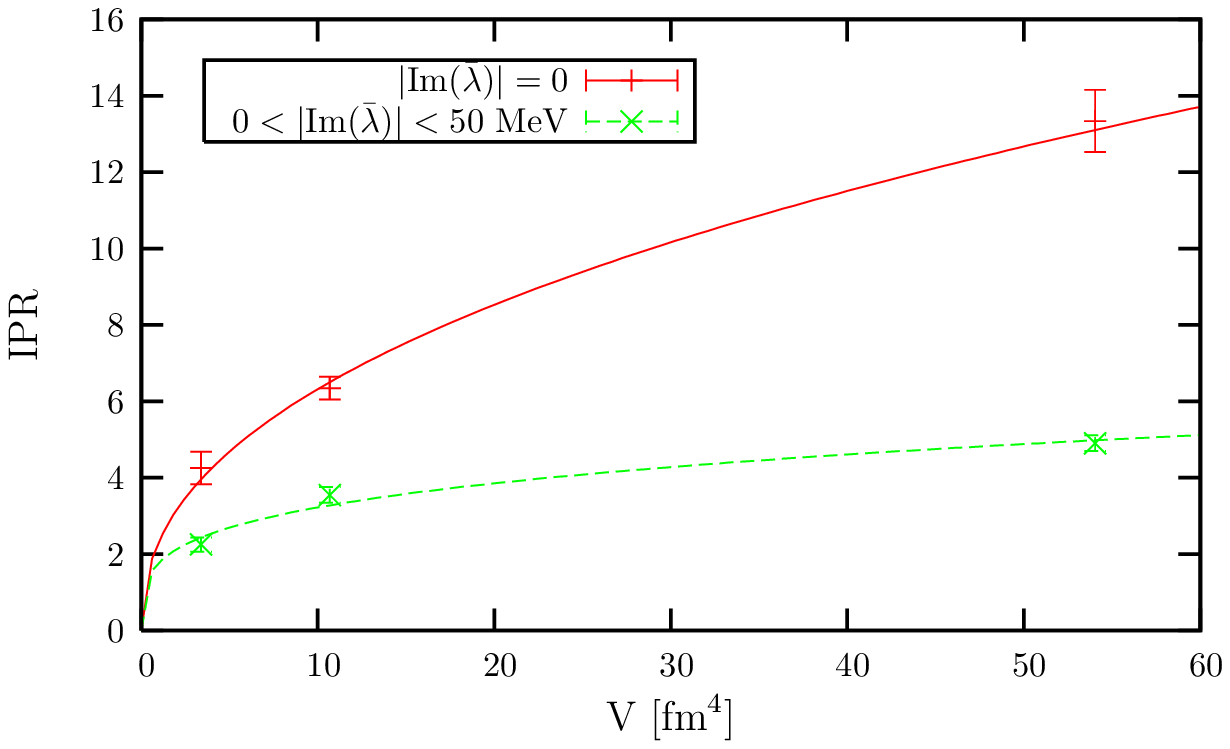,width=6cm}\\
(c) & (d)
\end{tabular}
\end{center}
\caption{The IPR of overlap eigenmodes (for all computed eigenmodes on all considered configurations) on the $12^3\times 24$ lattice (a) and the $24^3\times 48$ lattice (b) at $\beta=8.45$. The averaged IPR is shown in (c), while the volume dependence with fixed  $\beta=8.45$  for the zero and the low lying modes is shown in (d).}
\label{fig:ipr}
\end{figure*}

Fig.~\ref{fig:ipr}~(a) and (b) show the IPR for all eigenmodes on all considered  configurations as a function of the imaginary part of the eigenvalue on the $12^3\times 24$ and $16^3\times 32$ lattice at $\beta=8.45$. One can see that the zero modes and the first nonvanishing modes are, especially on the large lattice, highly localised, while the bulk of the higher modes is nonlocalised with  $I \lesssim 2$. In Fig.~\ref{fig:ipr}~(c) we plot the IPR averaged over bins with
bin size $\Delta {\rm Im} \bar\lambda = 50$ MeV, except
for the zero modes which are considered separately. 

From the volume dependence of the IPR one can derive the effective dimension $d$ of the eigenmodes based on the relation  $I \propto V^{1-d/4}$. Fig.~\ref{fig:ipr}~(d) shows the volume dependence of the averaged IPR at fixed lattice spacing. We fit the zero modes  and lowest nonzero modes with
 $0< | {\rm Im}~\bar\lambda | \le 50$ MeV separately and find $d$= 2.3(1) and $d$=3.0(2), respectively.
 
While the low modes strongly depend on $V$ and  $a$ and are thus localised with an effective dimension $d=2\ldots 3$, the bulk of the higher modes is independent on $V$ and $a$, and extends in $d=4$ dimensions.

\section{The local structure of topological charge fluctuations}

The topological charge density for any $\gamma_5$-Hermitean Dirac operator 
satisfying the Ginsparg-Wilson relation is defined as \cite{Hasenfratz:1998ri}:
\begin{equation}
q(x)=\frac{1}{2}\Tr\;\gamma_5\,D(x,x), \;\;Q=\sum_x q(x).
\label{eq-q}
\end{equation}

To compute the topological charge density, we use two different approaches \cite{Horvath:2002yn,Horvath:2003yj}.
In the first approach, we directly calculate the trace of the overlap operator 
according to Eq.~(\ref{eq-q}).
This is a computationally very demanding task and is therefore performed on only 53 (5) configurations on the $12^3\times 24$ ($16^3\times 32$) lattice so far.

\newcounter{limitedstat}
\setcounter{limitedstat}{\value{footnote}}
The full density computed in this way includes charge fluctuations at 
all scales. 

The second technique involves the  computation of the topological charge density based 
on the low lying modes of the overlap Dirac operator. 
This approach is gauge invariant and
leaves the lattice scale unchanged, in contrast to the cooling method.
Using the spectral representation of the Dirac operator, the truncated eigenmode expansion
of the topological charge density reads

\begin{eqnarray}
q_{\lambda_{\rm cut}}(x)&=&-\sum_{|\lambda|<\lambda_{\rm cut}}(1-\frac{\lambda}{2})\; c^\lambda(x),\\
c^{\lambda}(x)&=&\psi^\dagger_\lambda(x) \gamma_5 \psi_\lambda(x)\;.\nonumber
\end{eqnarray}
Truncating the expansion acts as an ultraviolet filter by removing the short-distance fluctuations from $q(x)$. Note that the topological charge $Q=\sum_x q_{\lambda_{\rm cut}}(x)$ is not affected by the level of truncation.

\begin{figure*}[t]
\begin{center}
\epsfig{file=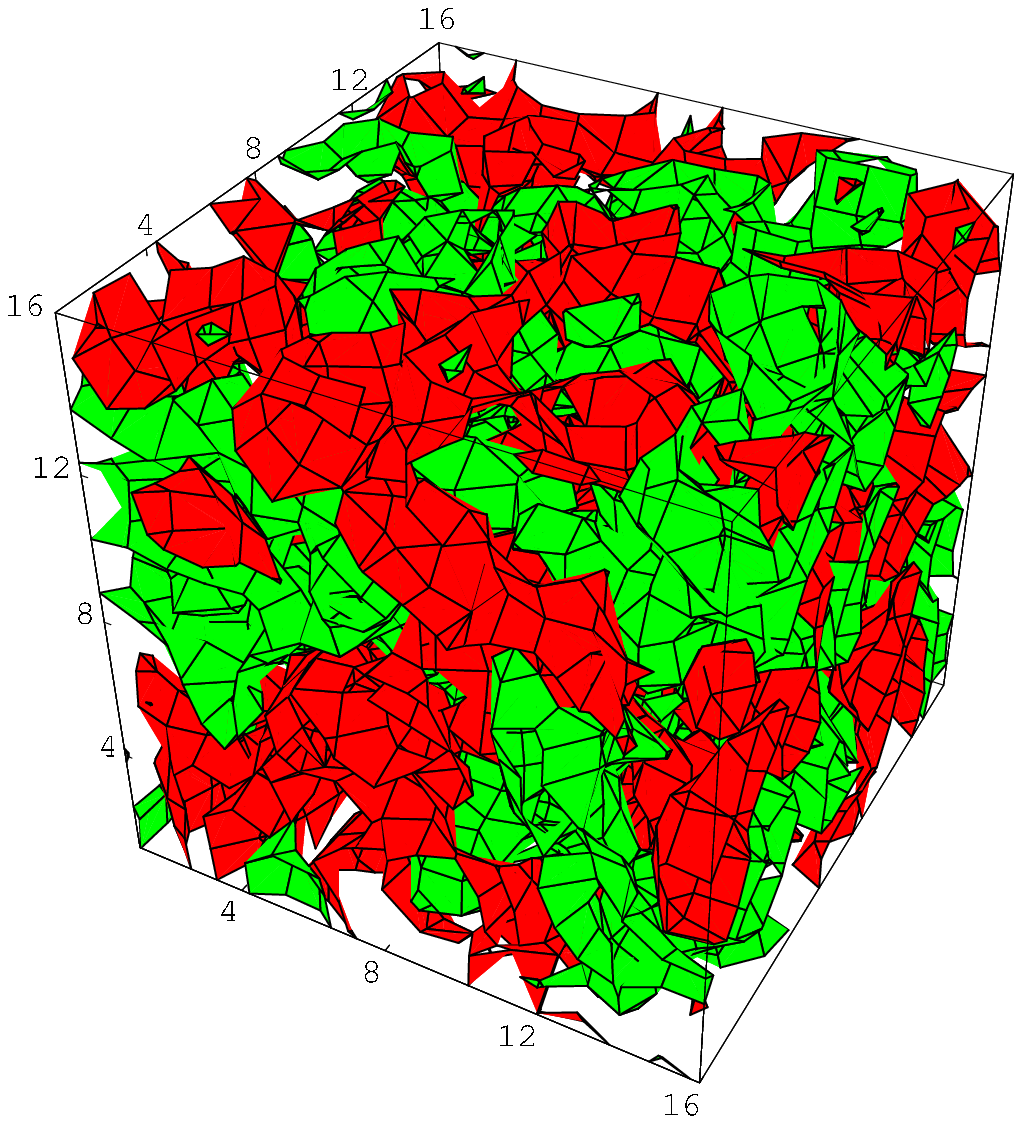,width=3.1cm}\hspace{0.3cm}   
\epsfig{file=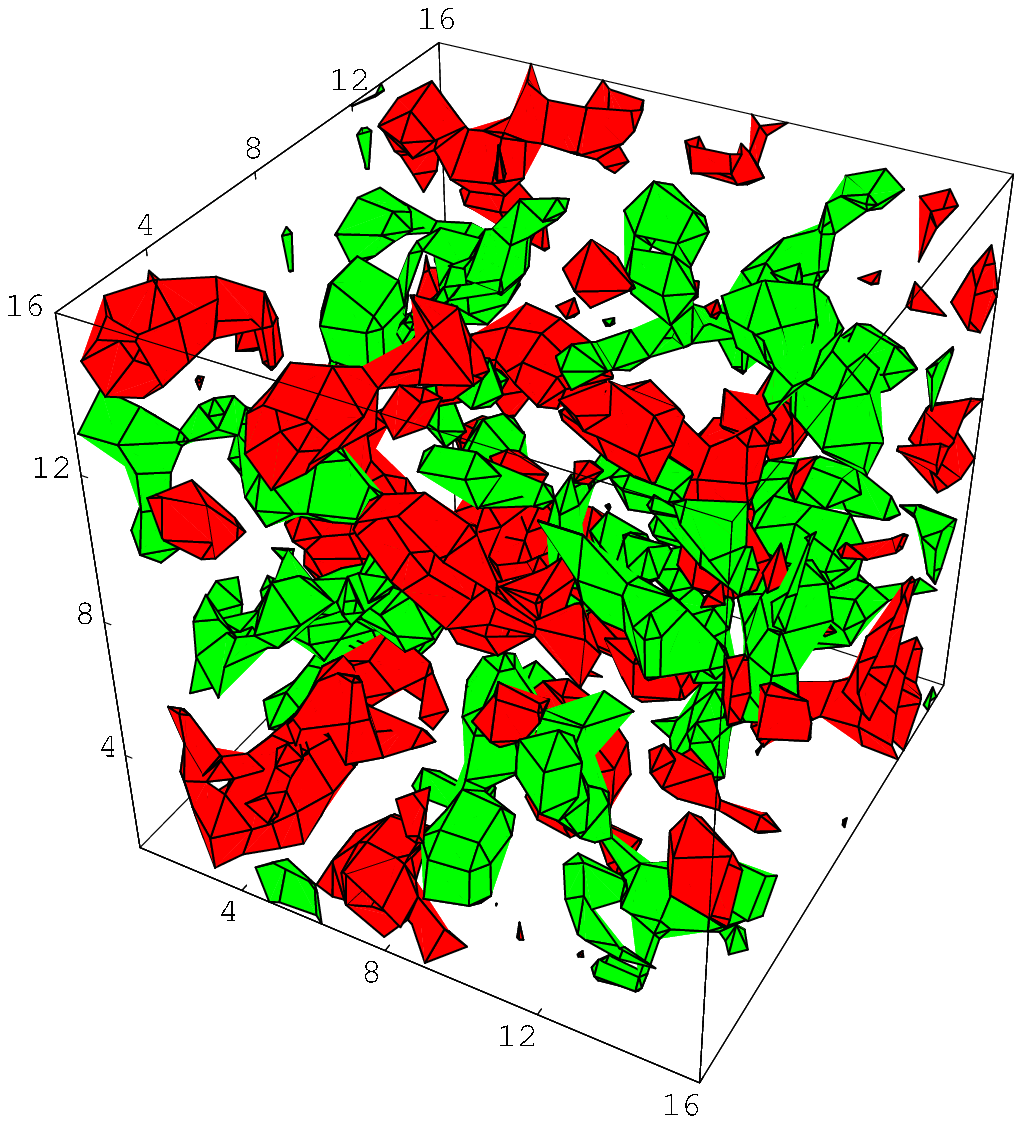,width=3.1cm}\hspace{0.3cm}   
\epsfig{file=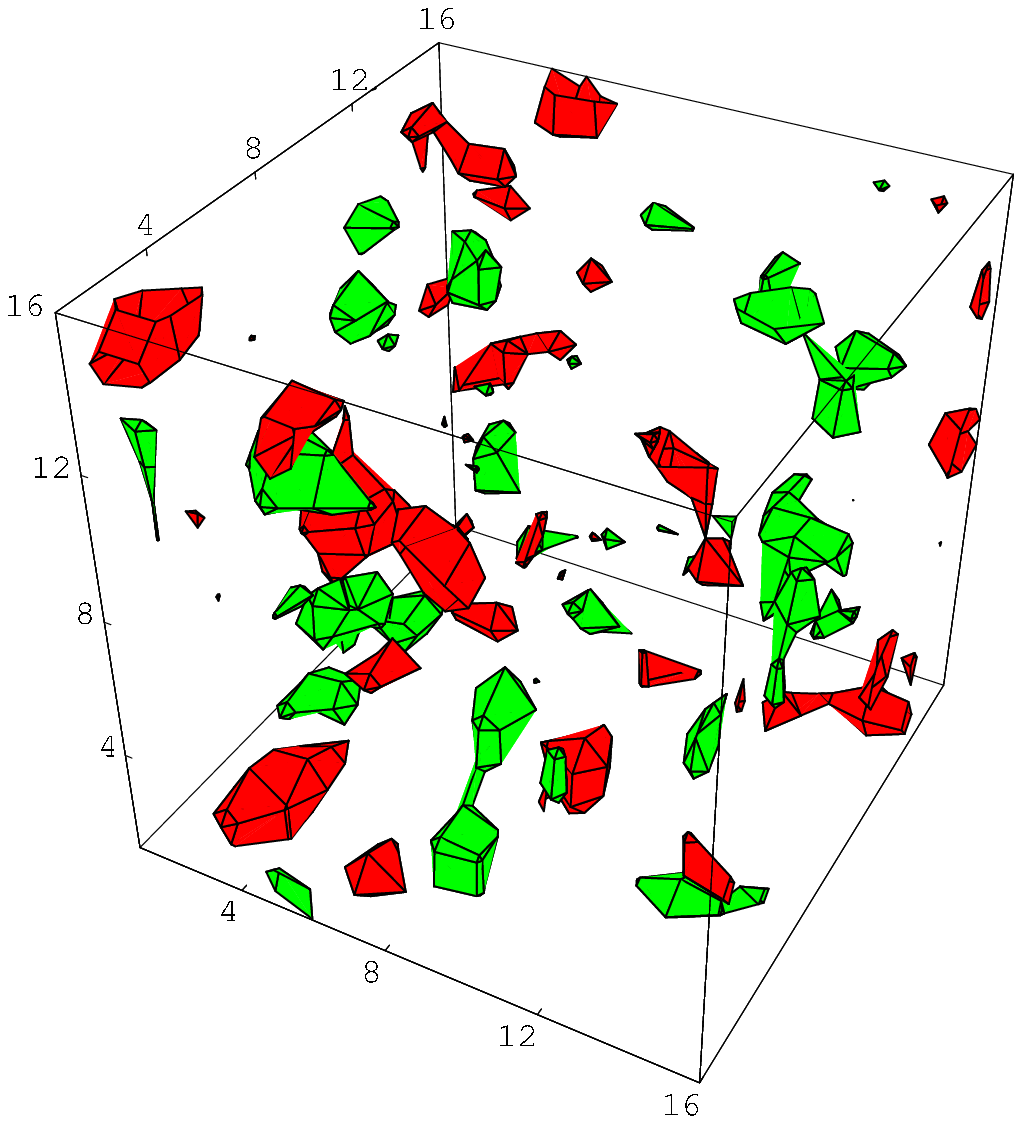,width=3.1cm}\hspace{0.3cm}   
\epsfig{file=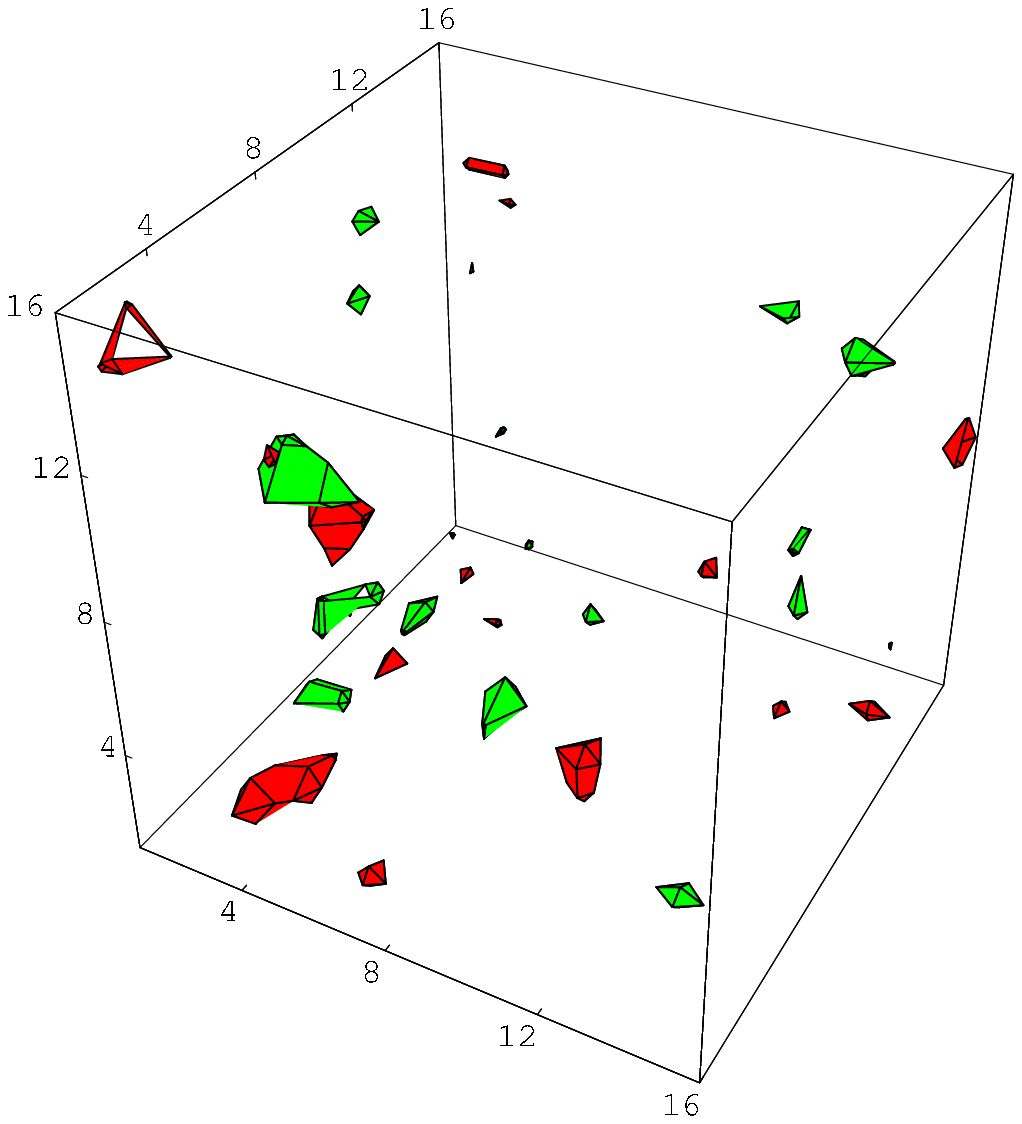,width=3.1cm}\hspace{0.3cm}\\
\epsfig{file=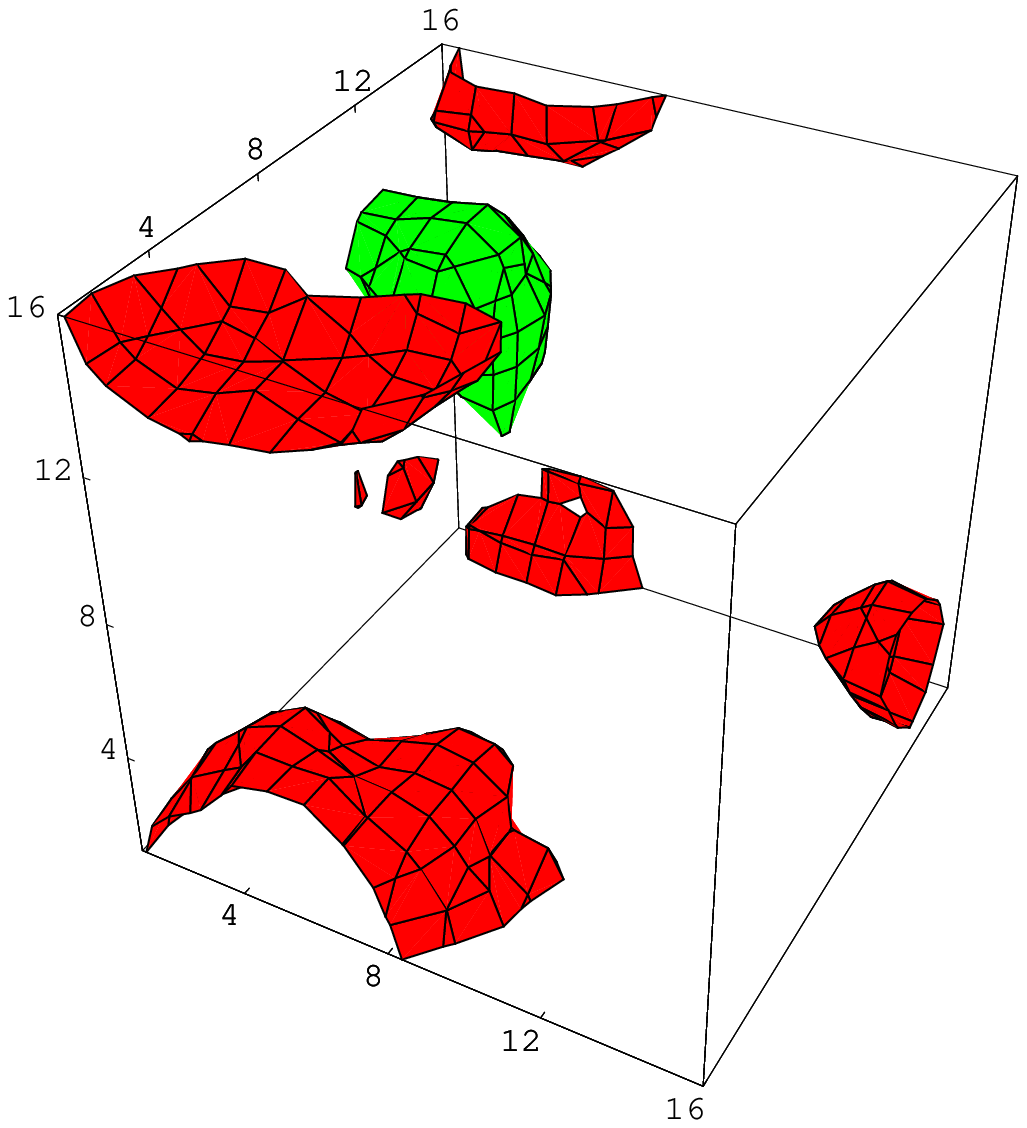,width=3.1cm}\hspace{0.3cm}   
\epsfig{file=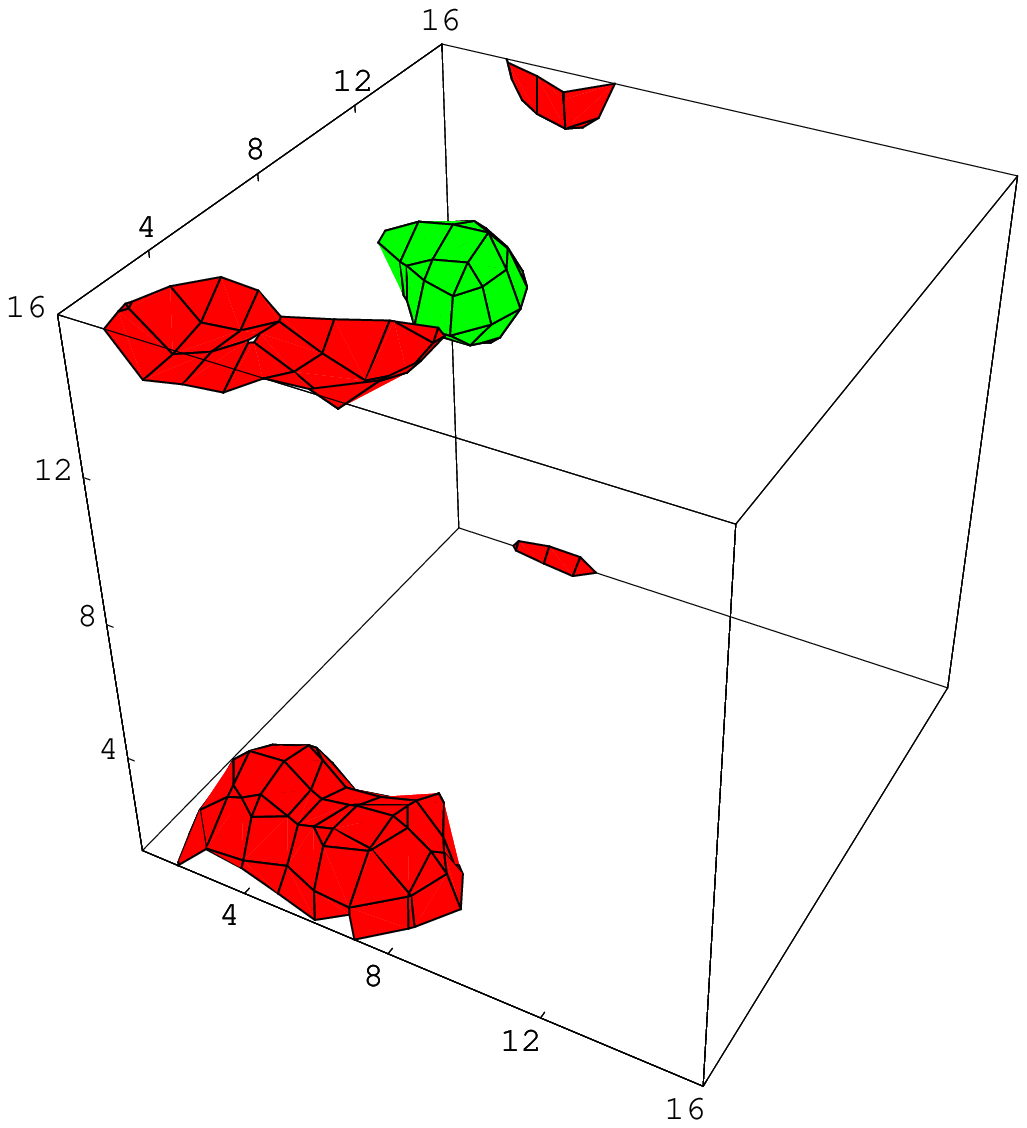,width=3.1cm}\hspace{0.3cm}   
\epsfig{file=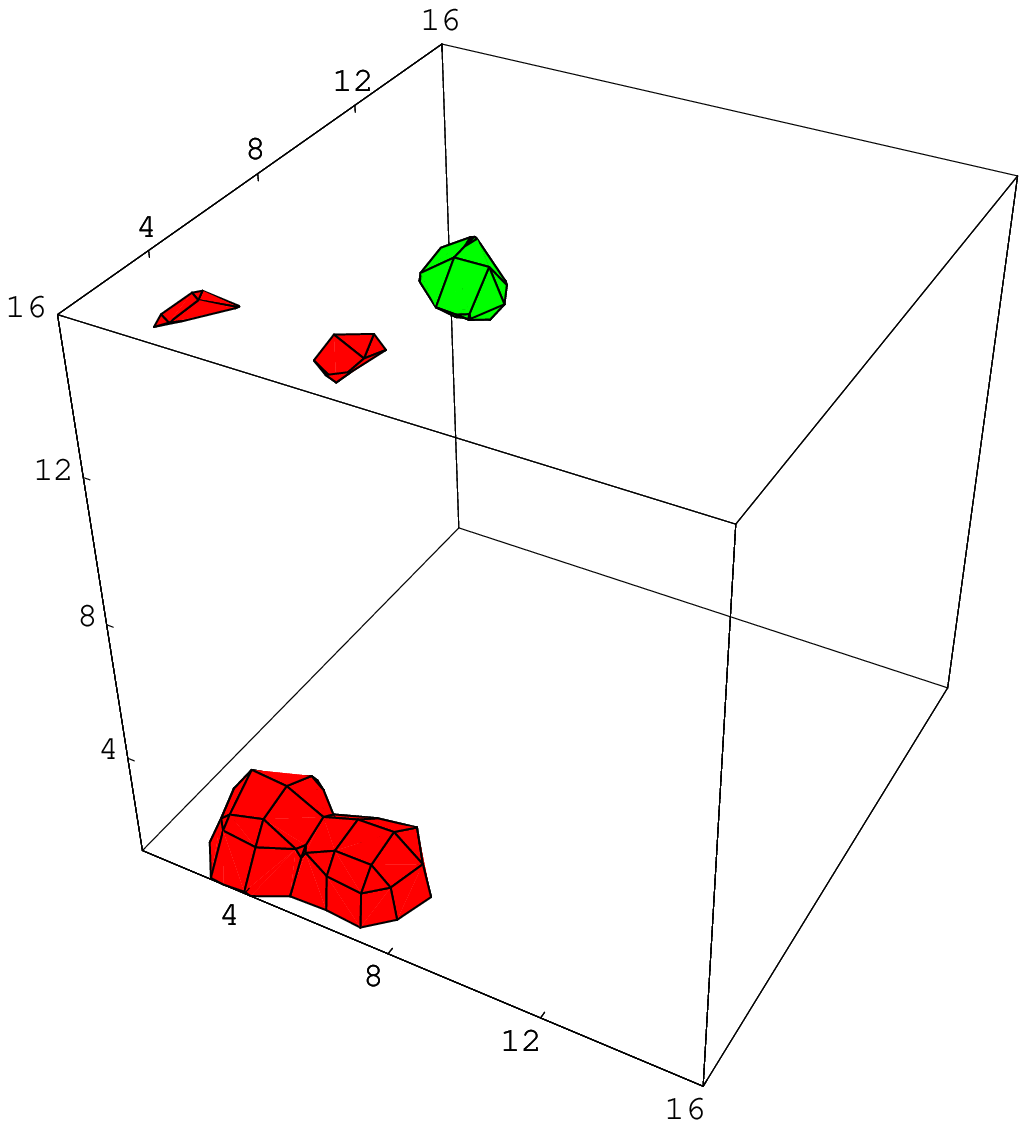,width=3.1cm}\hspace{0.3cm}   
\epsfig{file=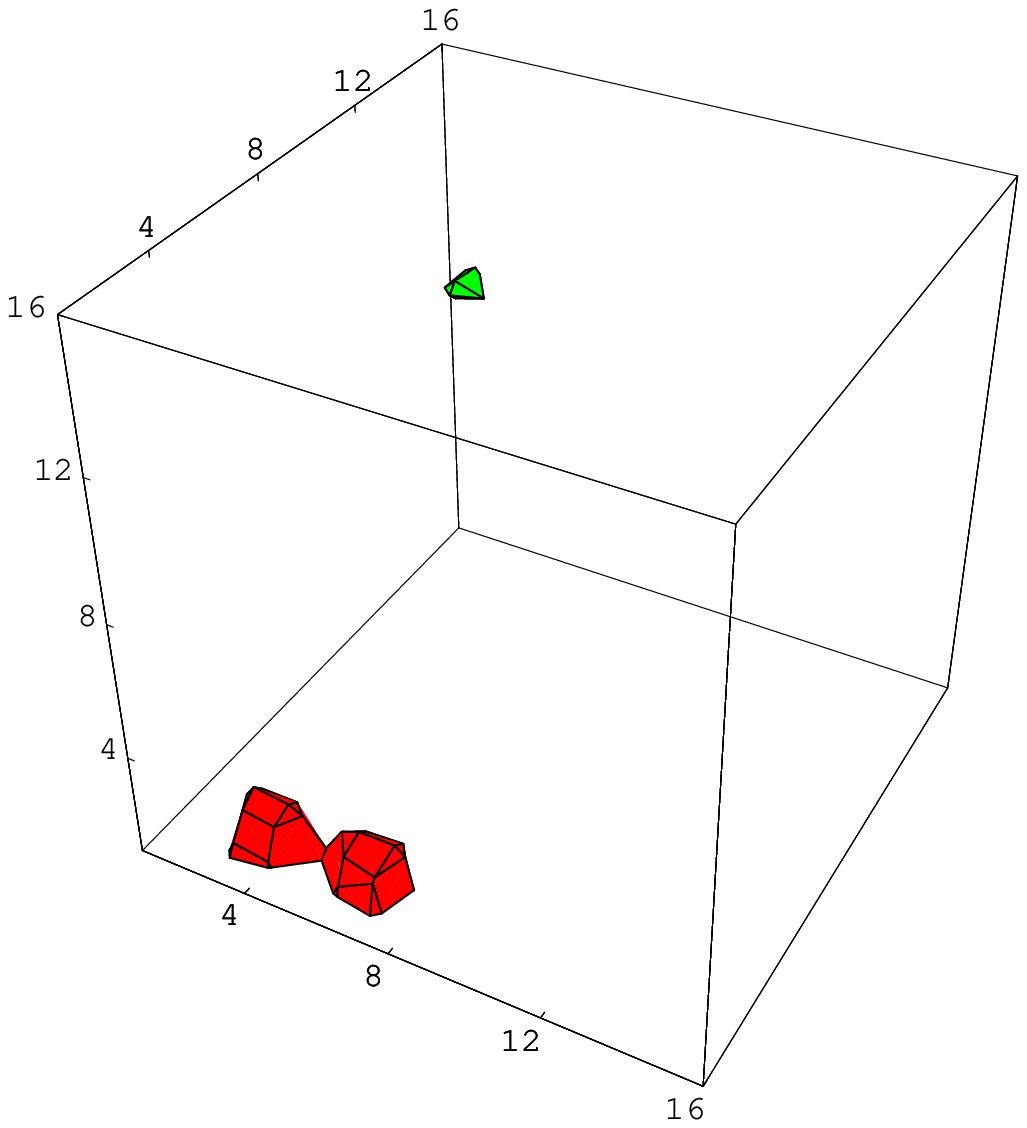,width=3.1cm}\hspace{0.3cm}   
\end{center}
\caption{Isosurfaces of topological charge density with 
$|q(x)|/q_{\rm max} = 0.1$, 0.2, 0.3, 0.4, on one timeslice of a single $16^3\times 32$ configuration at $\beta=8.45$.
The upper (lower) pictures are based on the full (eigenmode truncated, with $|\mbox{Im}(\bar\lambda)|\leq 0.076$) density. Colour encodes the sign of the surface. 
}
\label{fig-contour}
\end{figure*}

Let us first consider the local structure of the topological charge
fluctuations, by studying sign coherent clusters formed by connected 
neighbours $x_i$ with $|q(x_i)| > q_{\rm cut}  = (0.1 \dots 0.4)\: q_{\rm max} $, where $q_{\rm cut}$ is a 
cutoff value to be varied at will.
To give a picture of the spatial distribution of the topological charge density, isosurface plots with $q(x)=\pm q_{\rm cut}$ are shown in Fig.~\ref{fig-contour}.
It is obvious that the UV cutoff in the topological density results
in a  different topological structure. This is the structure actually seen by the lowest modes.

Reflection positivity, i.e. positivity of the metrics in Hilbert space, 
demands that the topological charge correlator is negative \cite{Seiler:2001je}

\pagebreak

\begin{equation}
\label{eq-seiler}
C_q(r)\!=\! \frac{1}{V} \sum_x  \langle q(x)q(y)\rangle \le 0, \; r\!=\!|x-y|>0 \:,
\end{equation}

\noindent where $r$ is the Euclidean distance. Since overlap fermions are not ultralocal, the fermion action is not
strictly reflection positive. Therefore 
one may expect a positive core and a negative tail of $C_q(r)$~\cite{Horvath:2005cv,Koma:2005}.

\begin{figure*}[h]
\centering
\hspace*{0.5cm}\epsfig{file=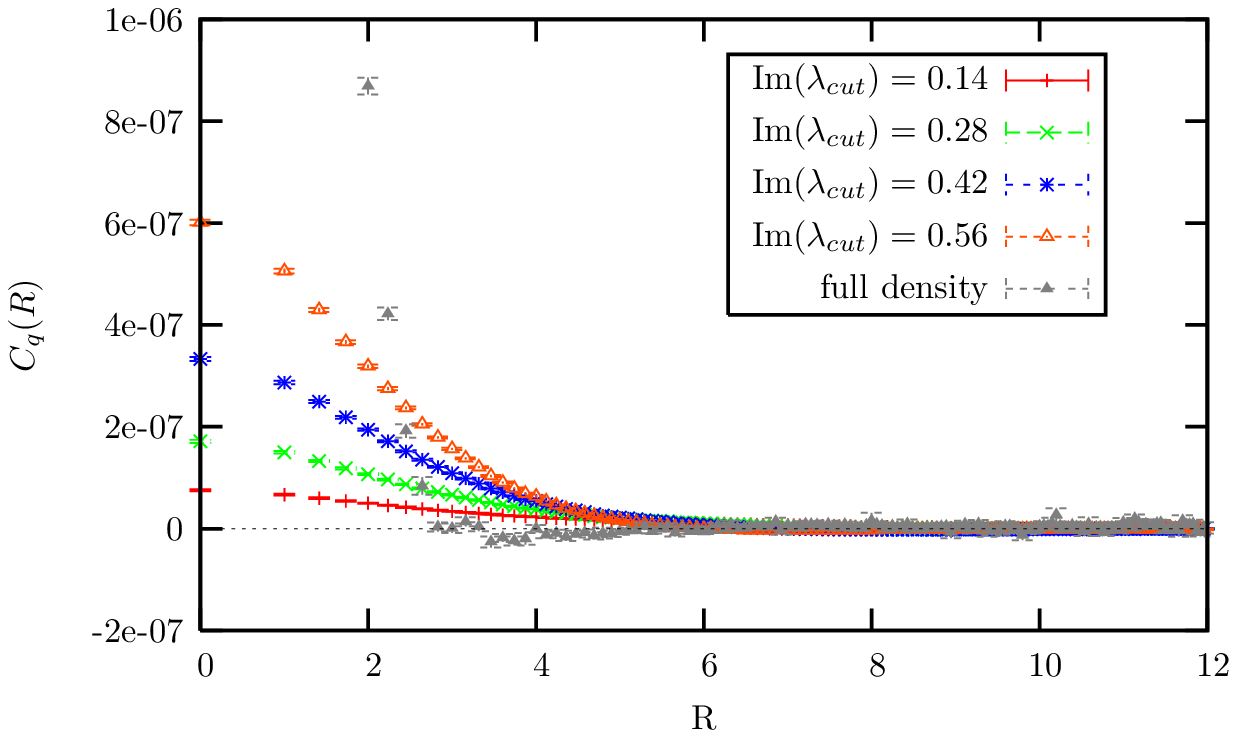,width=7.5cm}
\epsfig{file=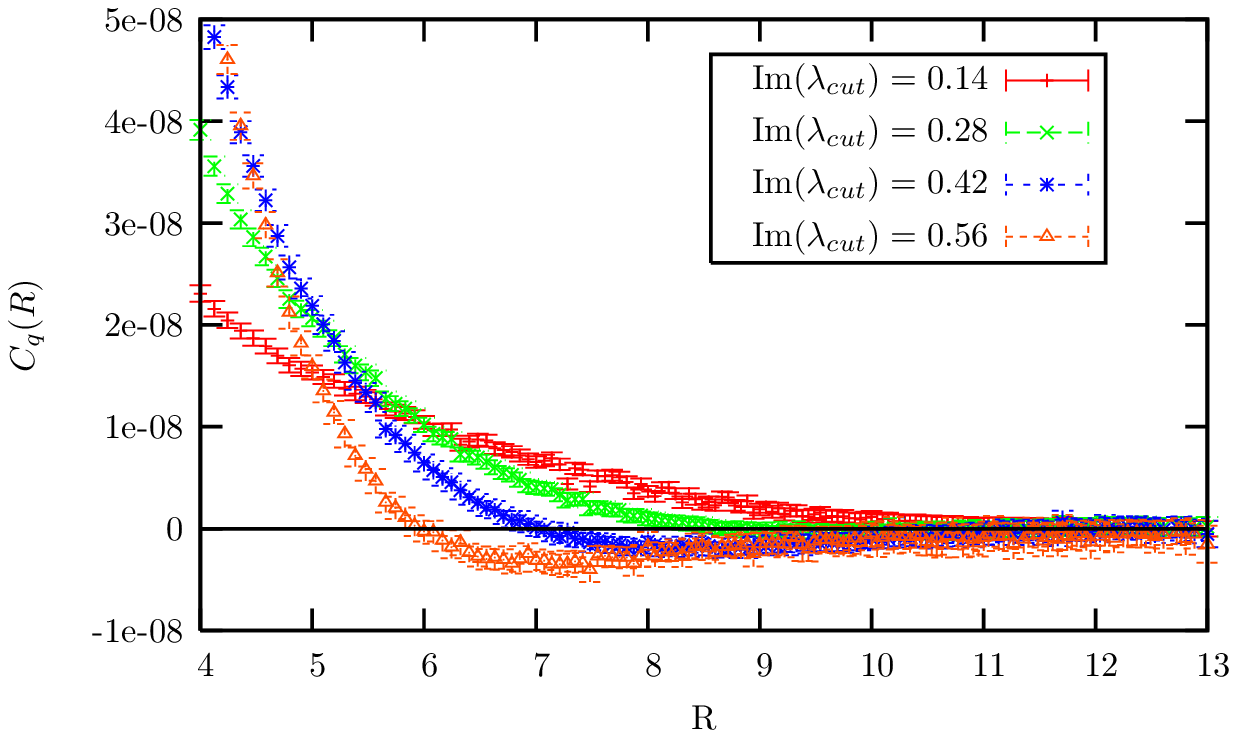,width=7.5cm}
\caption{The topological charge correlator
on the $12^{3}\times 24$ lattice at $\beta=8.10$.
We compare the correlator based on the eigenmode truncated version of $q(x)$ for various $\lambda_{\rm cut}$ with the one based on the full density.
The right figure magnifies the region in $R$ where the eigenmode truncated
correlator turns negative.}
\label{fig:qq1}
\vspace*{1cm}
\epsfig{file=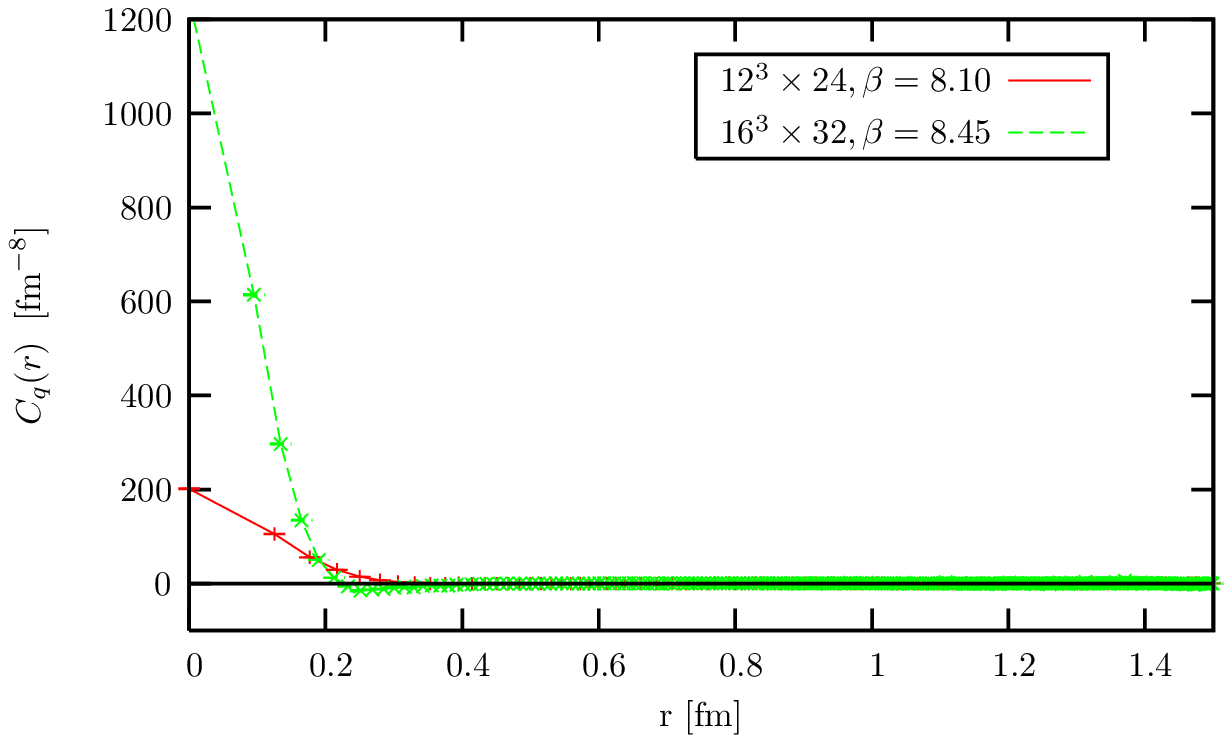,width=7.5cm}
\epsfig{file=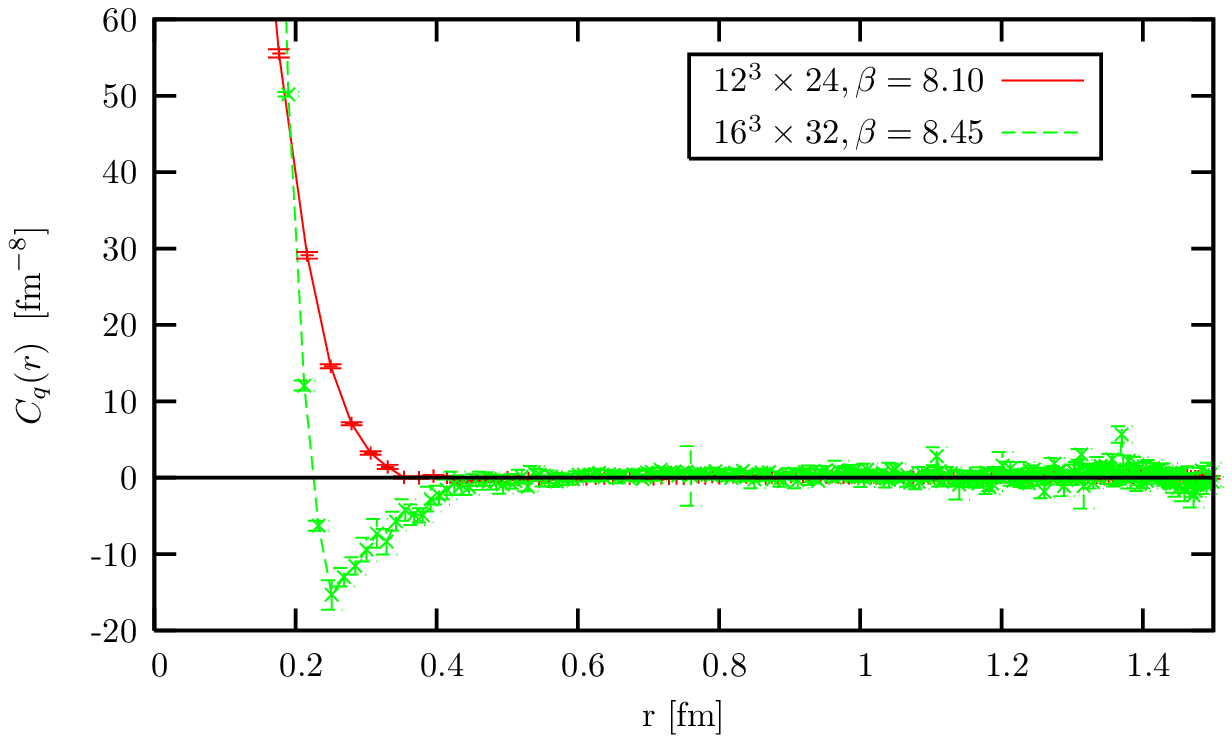,width=7.5cm}
\caption[]{The topological charge correlator based on the full charge density
on the $12^{3}\times 24$ lattice at $\beta=8.10$ ($N_{\rm conf}=53$) and
on the $16^{3}\times 32$ lattice at $\beta=8.45$ ($N_{\rm conf}=5$).
The right figure is a magnification of the left to show the negativity of the correlator.}
\label{fig:qq2}
\end{figure*}

Fig.~\ref{fig:qq1} shows that there is always a positive core near the origin
at any value of the cutoff, while the tail is only negative for sufficiently large cutoffs. Since the topological susceptibility is not affected by the cutoff in the number of modes, the growth of the positive core and the negative tail must compensate each other.

While the topological charge correlator $C_q(x)$ in the continuum should be negative for nonvanishing $x$, the topological susceptibility, which is just the integral over the correlator $\chi_t=\int dx\left<q(0)q(x)\right>=\int dx\,C_q(x)>0$,  must be positive. To solve the clash, formally divergent contact counterterms of the form  $c_1\delta(x)+c_2\Delta\delta(x)+c_3\Delta^2\delta(x)$  have to be introduced in the continuum theory \cite{Seiler:2001je}.

To see whether the expected behaviour is reached in the continuum limit, we examine in Fig.~\ref{fig:qq2} the $a$ dependence of the charge correlator based on the full density.
One can clearly see that the peak value of the positive core is increasing with decreasing $a$, while the support of the positive core is shrinking.
However, the finite  positive core at our highest $\beta$ remains at a
radius $R \approx 2\,a$, 
indicating the limits of locality of the operator $q(x)$ itself.

In the following we want to further investigate the cluster properties of the full density and compare our results obtained on the coarse $12^3 \times 24$ lattice at  $\beta=8.10$ and the finer $16^3\times 32$ lattice at $\beta=8.45$, which have a similar physical volume.

\begin{figure*}[h]
\begin{center}
\begin{tabular}{ccc}
\epsfig{file=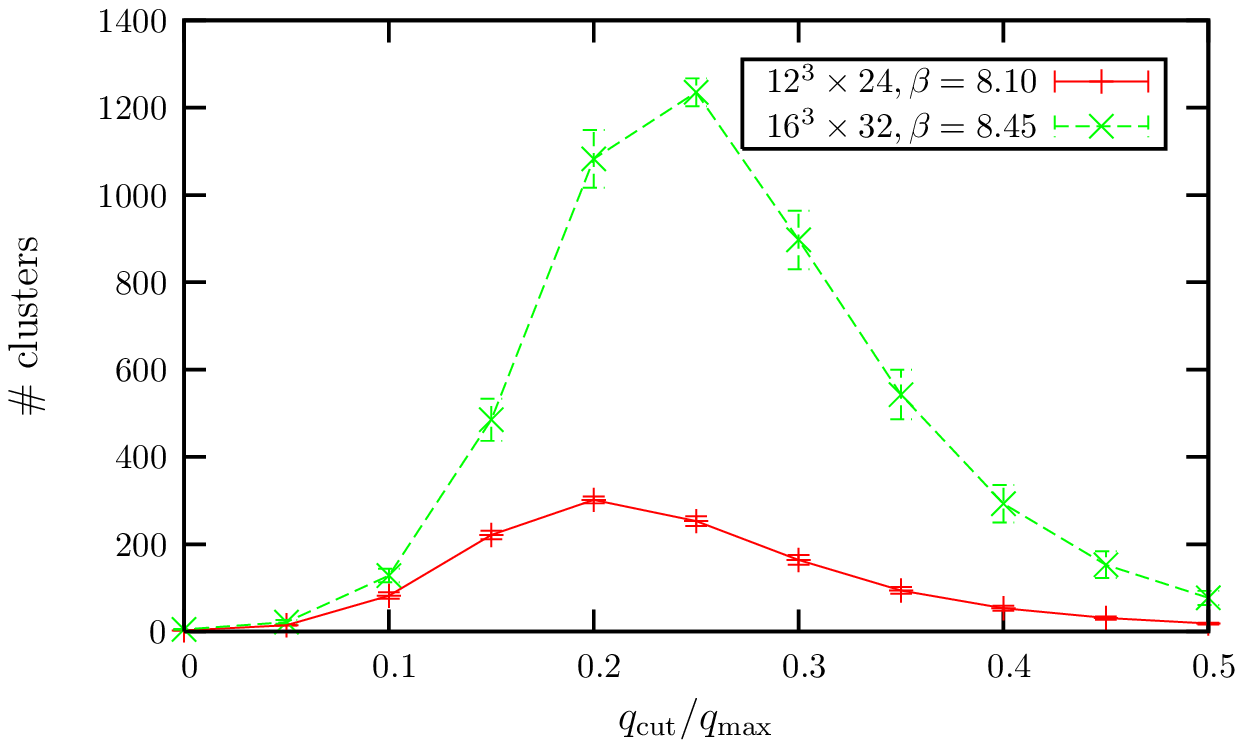, width=6cm}&
\epsfig{file=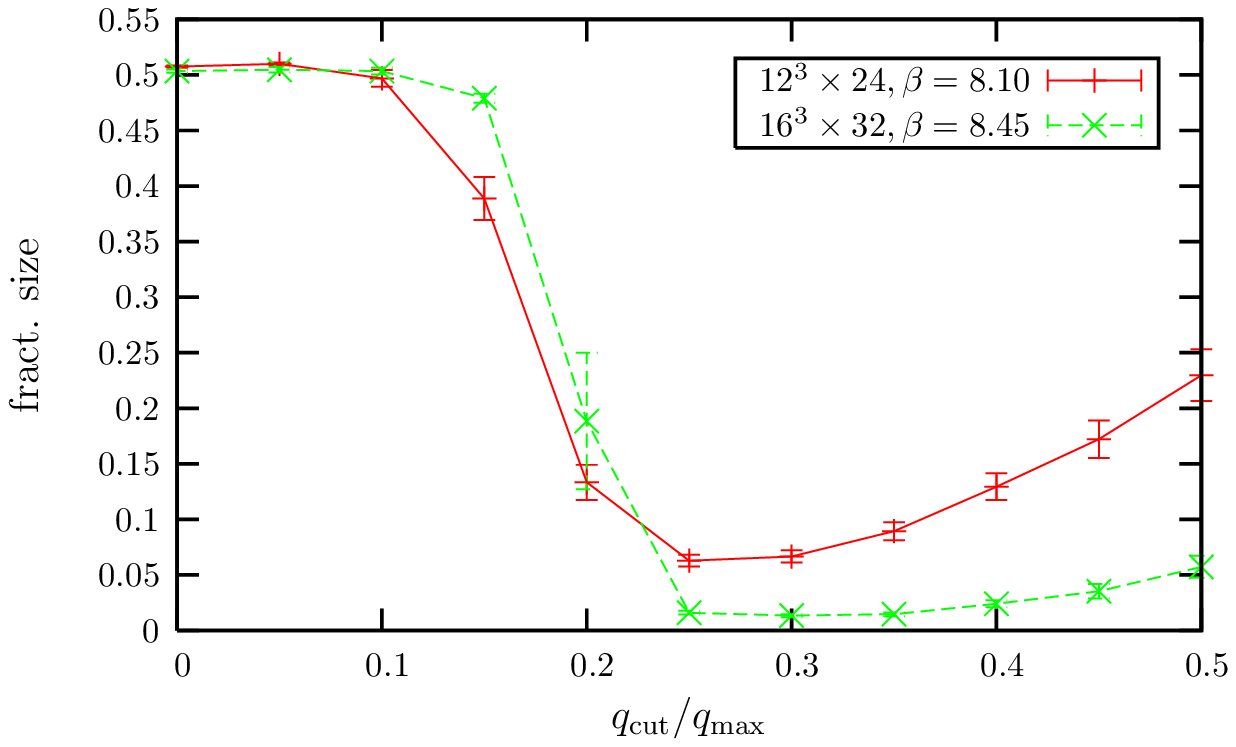, width=6cm} \\
(a) & (b)\\
\hspace*{0.3cm}\epsfig{file=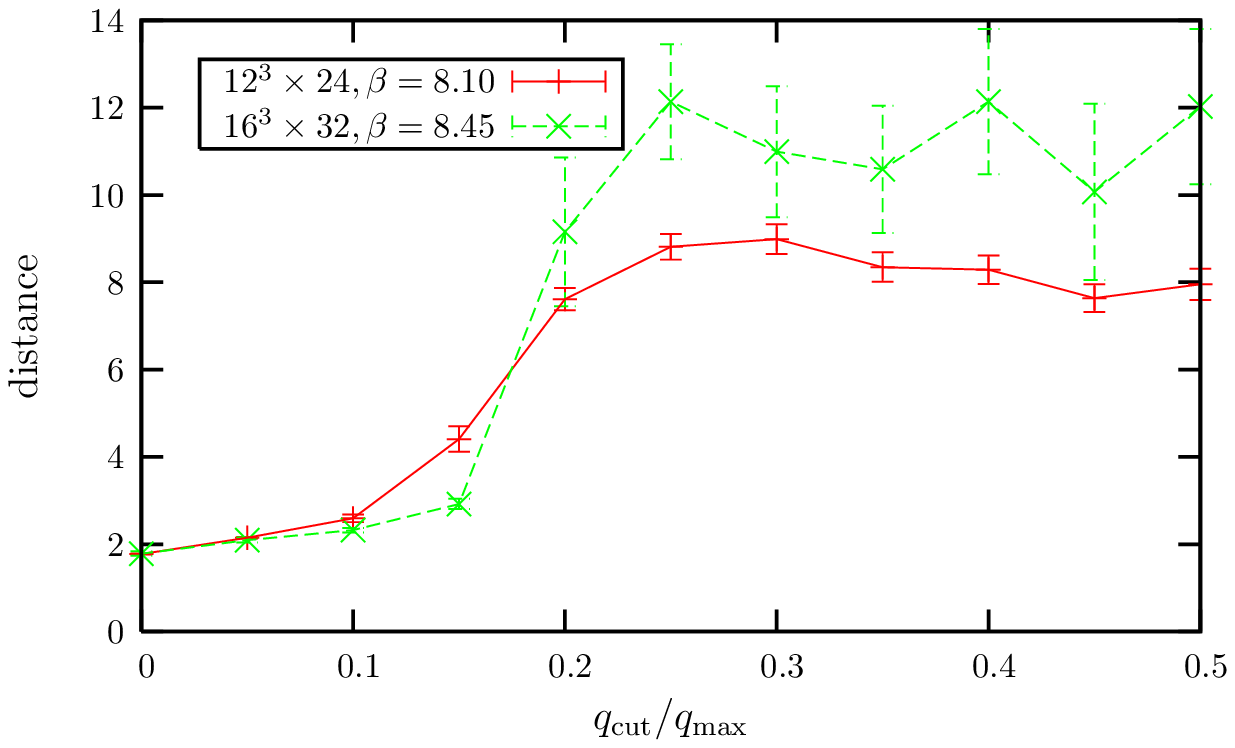, width=5.7cm}&
\epsfig{file=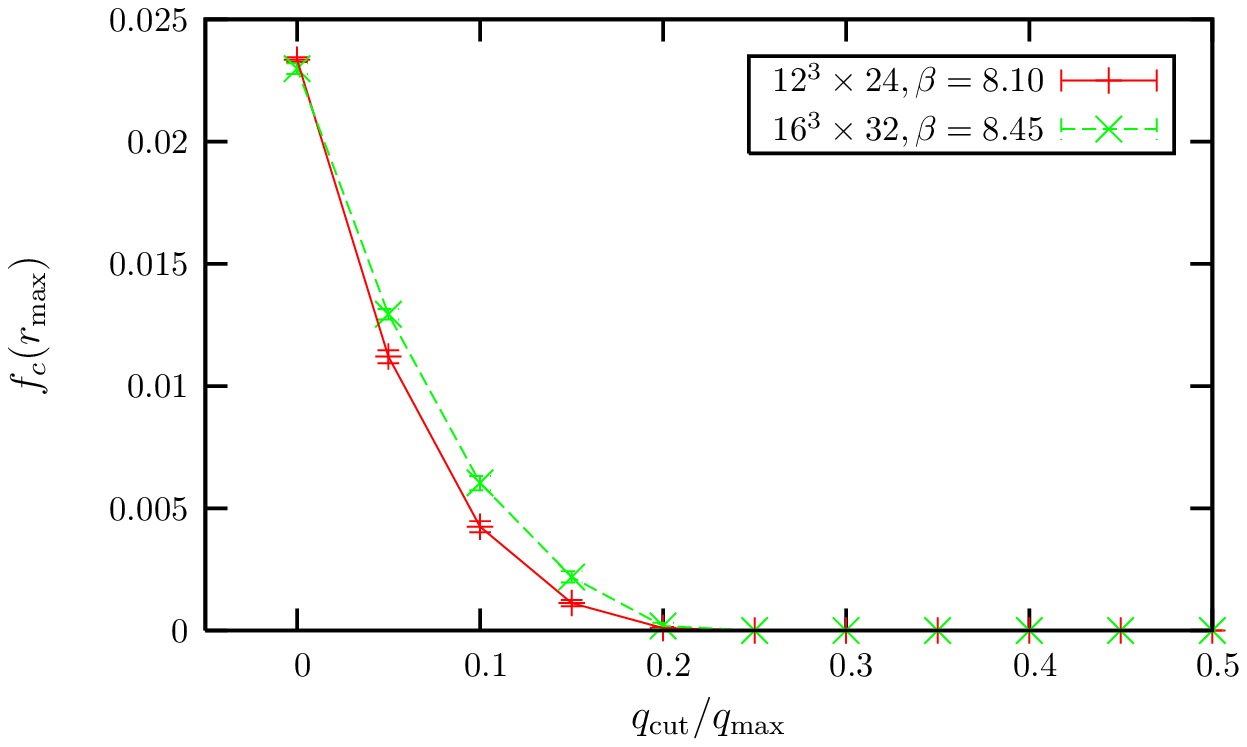, width=6cm}\\
(c) & (d)\\
\hspace*{-0.4cm}\epsfig{file=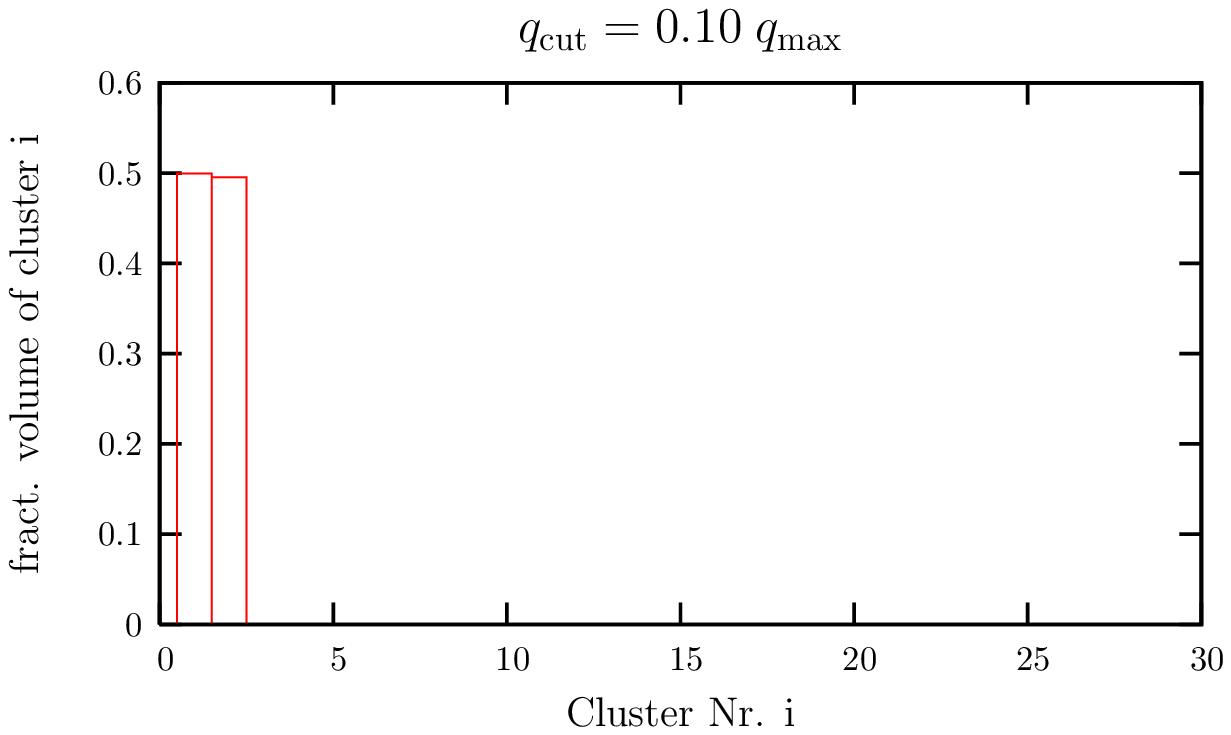,width=6.4cm}&
\hspace*{-0.8cm}\epsfig{file=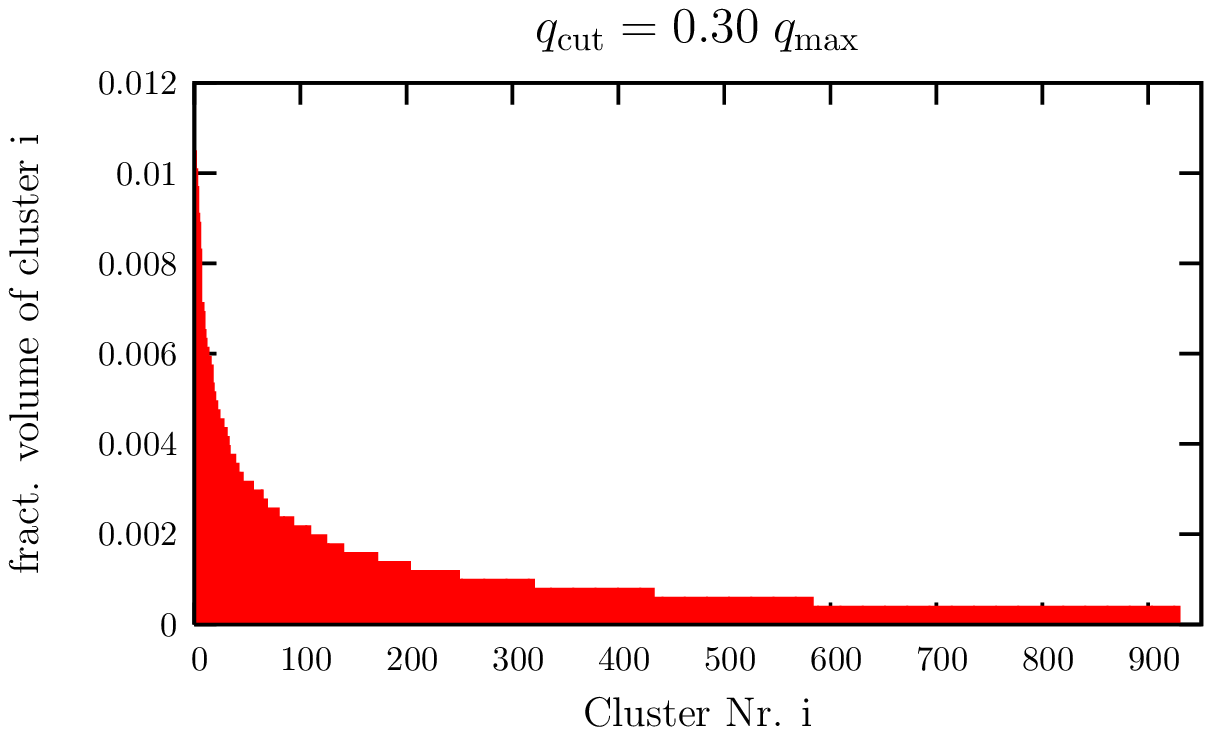,width=6.5cm}\\
(e) & (f)\\
\hspace*{-0.4cm}\epsfig{file=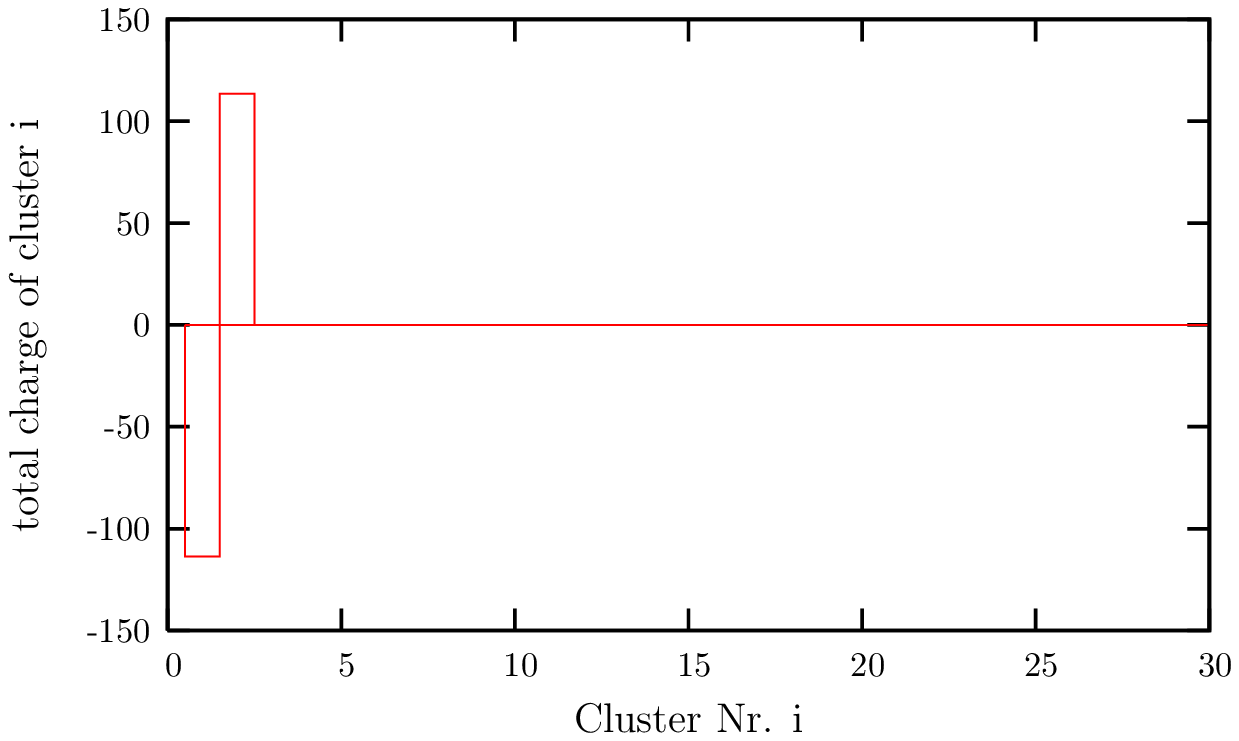,width=6.4cm}&
\hspace*{-0.5cm}\epsfig{file=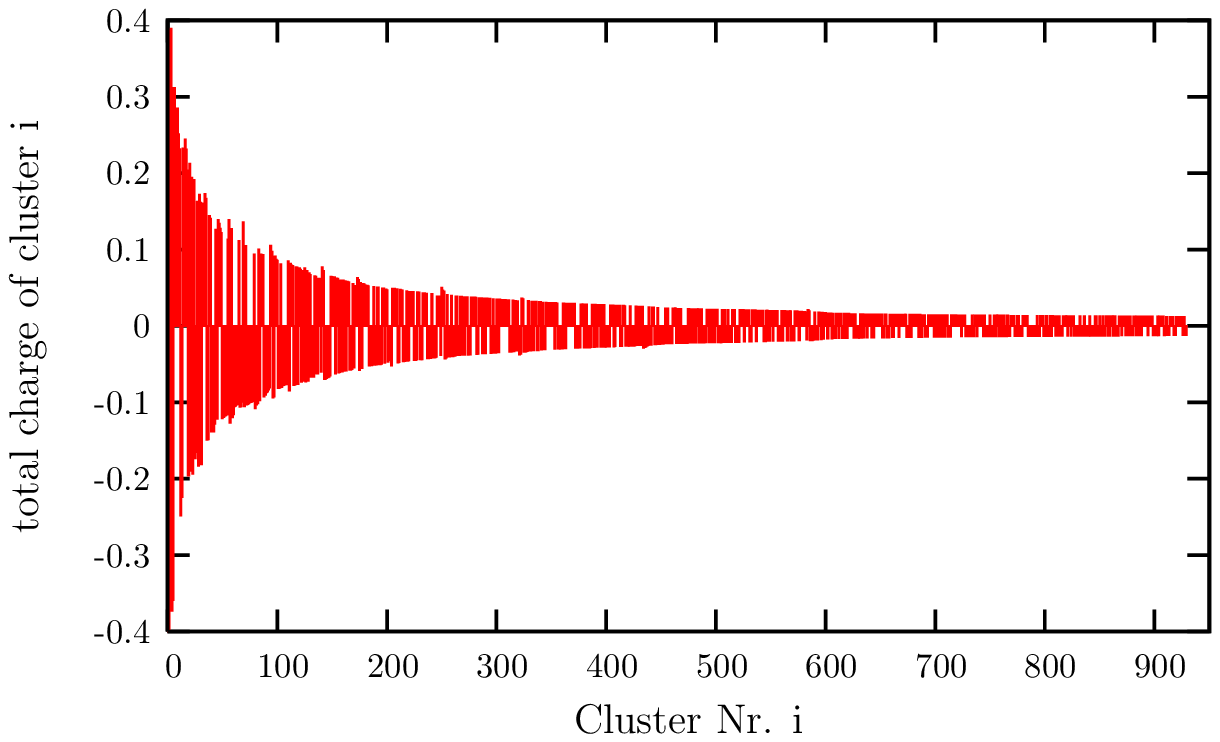,width=6.2cm}\\
(g) & (h)\\
\end{tabular}
\end{center}
\caption[]{Cluster analysis: The $q_{\rm cut}$ dependence of the total number of clusters of full density (a), the fractional size of the largest cluster (b), the distance between the 2 leading clusters (c)  and the cluster correlation function $f_c(R_{\rm max})$ at the largest possible distance (d) for the full densities. The data is plotted both for the $12^3\times 24$ lattice at $\beta=8.10$ and   
for the $16^3\times 32$ lattice at $\beta=8.45$, which have a similar physical volume.
 Also shown is the fractional volume (e)+(f) and the total charge (g)+(h) of the clusters of full charge sorted by their size for $q_{\rm cut}=0.10\; q_{\rm max}$ (left) and $0.30\; q_{\rm max}$  (right) for one single configuration on the $16^3\times 32$ lattice.}

\label{fig-cluster}
\end{figure*}

In Fig.~\ref{fig-cluster}~(a) we show that the  number of clusters\footnote{The maximal number of clusters of the truncated density is an order of magnitude smaller.}  reaches a maximum, 
which is higher nearer to the continuum, at 
$q_{\rm cut}$=0.20 (0.25) on the coarse (fine) lattice. For a cutoff as low as $q_{\rm cut}=0.05\; q_{\rm max}$
there are, independent of the lattice spacing, 2 totally dominating oppositely charged clusters \cite{Horvath:2003yj}.
Fig.~\ref{fig-cluster}~(b) demonstrates how the largest cluster approaches a packing fraction of 50~\%. The behaviour for the second largest cluster is similar. 

Fig.~\ref{fig-cluster}~(c) shows the maximum of the minimal distance between each point of the largest  cluster and every point of the second largest one. 
The small value for low cutoffs in $q$ implies that the 2 dominating clusters 
are tangled and intertwined in a complicated way.

\afterpage{\clearpage}

In Fig.~\ref{fig-cluster}~(d) we plot the point-to-point cluster correlation function $f_c(R)$, which measures the probability of a pair of points at a distance $R$ to be  in the same cluster, evaluated at the  largest  possible distance $R_{\rm max}$.  
We find that the largest cluster starts to percolate at $q_{\rm cut}\approx 0.20\; q_{\rm max}$ for the full density independent of the lattice spacing.

The fractional volume and the total charge of 
all clusters ordered by their size  on one typical configuration on the $16^3\times 32$ lattice are plotted in Figs.~\ref{fig-cluster}~(e)-(h).
Comparing the data for $q_{\rm cut}$=0.10 $q_{\rm max}$ and  0.30 $q_{\rm max}$ the striking difference between the percolating and nonpercolating regime can be observed.

A preliminary dimensional analysis indicates that we are rediscovering for the full density the picture of 2 dominating approximately 
3D sign coherent clusters \cite{Horvath:2005rv} 
at low cutoffs, whereas we see 1D highly charged small clusters at high 
thresholds. 
This picture means that the QCD vacuum model with 4D coherent
(anti)instantons with a typical instanton radius of $0.3-0.4$ fm
is strongly modified if quantum fluctuations of all scales are taken into consideration. The eigenmode-truncated density is more 
compatible with the conventional multiple-lump picture.

\section*{Acknowledgements}

The numerical calculations have been performed at NIC J\"ulich, HLRN Berlin, DESY-Zeuthen, LRZ Munich and the CIP Physik pool at the University of Munich. 

We thank these institutions for support. Part of this work is 
supported by DFG under contract FOR 465 
(Forschergruppe Gitter-Hadronen Ph\"anomenologie).

\bibliographystyle{aip}
\bibliography{proc}

\begin{thebibliography}{10}

\bibitem{Neuberger:1997fp}
H.~Neuberger,
\newblock Phys. Lett. {\bf B417}, 141 (1998).

\bibitem{Neuberger:1998wv}
H.~Neuberger,
\newblock Phys. Lett. {\bf B427}, 353 (1998).

\bibitem{Luscher:1998pq}
M.~L\"uscher,
\newblock Phys. Lett. {\bf B428}, 342 (1998).

\bibitem{Capitani:1999uz}
S.~Capitani, M.~G\"ockeler, R.~Horsley, P.~E.~L. Rakow, and G.~Schierholz,
\newblock Phys. Lett. {\bf B468}, 150 (1999).

\bibitem{Giusti:2002sm}
L.~Giusti, C.~H\"olbling, M.~L\"uscher, and H.~Wittig,
\newblock Comput. Phys. Commun. {\bf 153}, 31 (2003).

\bibitem{Gockeler:1989qg}
M.~G\"ockeler, A.~S. Kronfeld, M.~L. Laursen, G.~Schierholz, and U.~J. Wiese,
\newblock Phys. Lett. {\bf B233}, 192 (1989).

\bibitem{Luscher:1984xn}
M.~L\"uscher and P.~Weisz,
\newblock Commun. Math. Phys. {\bf 97}, 59 (1985).

\bibitem{Gattringer:2001jf}
C.~Gattringer, R.~Hoffmann, and S.~Schaefer,
\newblock Phys. Rev. {\bf D65}, 094503 (2002).

\bibitem{Banks:1979yr}
T.~Banks and A.~Casher,
\newblock Nucl. Phys. {\bf B169}, 103 (1980).


\bibitem{Chiu:1998gp}
T.~W.~Chiu and S.~V.~Zenkin,
\newblock Phys.\ Rev. {\bf D59}, 074501 (1999).

\bibitem{Chiu:1998aa}
T.~W.~Chiu, C.~W.~Wang and S.~V.~Zenkin,
\newblock  Phys.\ Lett. {\bf B438}, 321 (1998).

\bibitem{Osborn:1998qb}
J.~C. Osborn, D.~Toublan, and J.~J.~M. Verbaarschot,
\newblock Nucl. Phys. {\bf B540}, 317 (1999).

\bibitem{Damgaard:2001xr}
P.~H. Damgaard,
\newblock Nucl. Phys. {\bf B608}, 162 (2001).

\bibitem{Wilke:1997gf}
T.~Wilke, T.~Guhr, and T.~Wettig,
\newblock Phys. Rev. {\bf D57}, 6486 (1998).

\bibitem{Gattringer:2001mn}
C.~Gattringer, M.~Gockeler, P.~E.~L.~Rakow, S.~Schaefer and A.~Schafer,
\newblock Nucl.\ Phys. {\bf B617}, 101 (2001).

\bibitem{Aubin:2004mp}
C.~Aubin et~al.,
\newblock Nucl. Phys. Proc. Suppl. {\bf 129}, 626 (2004).


\bibitem{Gubarev:2005jm}
F.~V.~Gubarev, S.~M.~Morozov, M.~I.~Polikarpov and V.~I.~Zakharov,
\newblock JETP Lett. {\bf 82}, 343 (2005).


\bibitem{Hasenfratz:1998ri}
P.~Hasenfratz, V.~Laliena, and F.~Niedermayer,
\newblock Phys. Lett. {\bf B427}, 125 (1998).

\bibitem{Horvath:2002yn}
I.~Horvath et~al.,
\newblock Phys. Rev. {\bf D67}, 011501 (2003).

\bibitem{Horvath:2003yj}
I.~Horvath et~al.,
\newblock Phys. Rev. {\bf D68}, 114505 (2003).

\bibitem{Seiler:2001je}
E.~Seiler,
\newblock Phys. Lett. {\bf B525}, 355 (2002).

\bibitem{Horvath:2005cv}
I.~Horvath et~al.,
\newblock Phys. Lett. {\bf B617}, 49 (2005).

\bibitem{Koma:2005}
Y.~Koma et~al.,
\newblock PoS(LAT2005)300  (2005).

\bibitem{Horvath:2005rv}
I.~Horvath et~al.,
\newblock Phys. Lett. {\bf B612}, 21 (2005).

\end{thebibliography}

\end{document}